\documentclass[12pt]{article} 

\usepackage{mathbbol,amssymb,latexsym,amsfonts,amsmath,amsthm}

\def\pmx{\begin{pmatrix}}
\def\emx{\end{pmatrix}}
\def\bsq{\begin{subequations}}
\def\esq{\end{subequations}}
\def\be{\begin{eqnarray}}
\def\ee{\end{eqnarray}}
\def\bee{\begin{eqnarray*}}
\def\eee{\end{eqnarray*}}
\def\bal{\begin{align}}
\def\eal{\end{align}}

\def\ds{\displaystyle}

 \newtheorem{thm}{Theorem}

\newtheorem{cor}[thm]{Corollary}
\newtheorem{prob}{Problem}
\newtheorem{conj}[prob]{Conjecture}

\def\id{{\cal I}}

\def\bra{\langle}
\def\ket{\rangle}

\def\kb{ \ket \bra }

\def\raw{\rightarrow}

\def\half{{\textstyle \frac{1}{2}}}

 \def\tr{\hbox{Tr} \,}
 \def\trp{\hbox{Tr} }
\def\mm{ \! - \! }
\def\pp{ \! +\! }

\def\nn{\nonumber}
\def\ot{\otimes}

\def\wtd{\widetilde}
\def\wh{\widehat}
\def\ovb{\overline}
\def\td{\tfrac{1}{d}}
\def\s2{\tfrac{s}{2}}
\def\e{\epsilon}

\def\mm{ \! - \! }
\def\pp{ \! + \! }

\def\qed{\qquad{\bf QED}}
\def\pf{ \noindent{\bf Proof:} }

\newcommand{\proj}[1]{ | #1 \kb  #1|}
\newcommand{\norm}[1]{ \| #1  \|}

\title{\bf \Large Open Problems in Quantum Information Theory }

\author{  Mary Beth Ruskai\thanks{Partially supported   
   by the National Science  Foundation under Grant  DMS-0604900} 
  \\  {\small Department of Mathematics,
     Tufts University,
       Medford, MA 02155} \\
     {\small     Marybeth.Ruskai@tufts.edu}}

\begin{document}

\maketitle

\begin{abstract}
Some open questions in quantum information theory (QIT) are described.
Most of them were presented in Banff during the BIRS workshop on Operator Structures  in QIT
11-16 February 2007.   New material has been added in view of the recent
counter-examples to p-norm multiplicativity.
\end{abstract}

\tableofcontents

\section{Extreme points of CPT maps}

In QIT, a channel is represented by a completely-positive
 trace-preserving (CPT) map 
$\Phi: M_{d_1}  \mapsto M_{d_2}$, which  is often written in the 
Choi-Kraus form 
\be  \label{CK}
\Phi(\rho) = \sum_k A_k \rho A_k^\dag \qquad  \text{with} \qquad 
 \sum_k A_k^\dag A_k = I_{d_1}.
 \ee
  The state representative or Choi matrix of $\Phi$ is
\be  \label{CJ}
    \Phi(\proj{\beta}) = \tfrac{1}{d} \sum_{jk} |e_j \kb e_k | \Phi( |e_j \kb e_k |)
\ee
where $|\beta \ket$ is a maximally entangled Bell state.  
Choi \cite{Choi} showed that the $A_k$ can be obtained from
the eigenvectors of  $ \Phi(\proj{\beta}) $ with non-zero eigenvalues.
The operators $A_k$ in \eqref{CK} are known to be defined only up to
a partial isometry and are often called Kraus operators.    When a minimal
set is obtained from Choi's prescription using eigenvectors of \eqref{CJ},
they are defined up to mixing of those from degenerate eigenvalues
 and we will refer to them as  Choi-Kraus operators.   
Choi showed that $\Phi$ is an extreme 
point of the set of CPT maps $\Phi: M_{d_1}  \mapsto M_{d_2}$ if and
only if the set  $\{ A_j^\dag A_k \}$ is linearly independent in $M_{d_1}$.
This implies that the Choi matrix of an extreme CPT map has rank
at most $d_1$.    We will refer to the rank of \eqref{CJ} as the {\em Choi rank}
of $\Phi$.   (Note that this is {\em not} the same as the rank of $\Phi$
as a linear operator from $M_{d_1}$ to $M_{d_2}$.)   

 It is often useful to consider the set of all CPT maps with 
Choi rank $\leq d_1$.   In \cite{RSW} these were called   ``generalized extreme points''
and shown to be equivalent to the closure of the set of extreme points
for qubit maps.    This is true in general.
Let ${\cal E}(d_1,d_2)$ denote the extreme points of the convex set of 
CPT maps from $M_{d_1}$ to $M_{d_2}$.  
\begin{thm}  The closure
$\ovb{{\cal E}(d_1,d_2)}$ of the set of extreme points of CPT maps
$\Phi: M_{d_1}  \mapsto M_{d_2}$ is precisely the set of such maps
with Choi rank  at most $d_1$.   
  \end{thm}
\pf  Let     Let $A_k$ be the Choi-Kraus operators for a
map $\Phi: M_{d_1}  \mapsto M_{d_2}$ with Choi rank $r \leq d_1$ which
is not extreme, and let $B_k$ be the Choi-Kraus operators for a true extreme point
with Choi-rank $d_1$.
When $r < d_1$ extend $A_k$ by letting $A_m = 0$ for $m = r \pp 1, r \pp 2, \ldots d_1$
and define $C_k(\e) = A_k + \e B_k$.     There is a number $\e_* $ such that 
the $d_1^2$ matrices  $C_j^\dag(\e) C_k(\e) $ are linear independent
for $0  <  \e < \e_* $.     To see this,  for each $C_j^\dag(\e) C_k(\e) $ ``stack'' 
the columns to give a  
vector of length $d_1^2$ and let  $M(\e)$ denote the  $d_1^2  \times d_1^2$
  matrix formed with these vectors as columns.   Then $\det M(\e)$ is a polynomial of
degree $d_1^4$, which has  at most  $d_1^4$ distinct roots.  Since the matrcies
$A_j^\dag A_k$ were assumed to be linearly dependent, one of these 
roots is $0$; it suffices to take  $\e_*$ the next largest root (or $+1$ if no 
roots are positive).  
Thus, the operators $C_j^\dag(\e) C_k(\e) $ are linearly independent  for $\e \in (0, \e_* )$.
The map  $\rho \mapsto \sum_k C_k(\e) \rho \, C_k^\dag(\e)$ is CP,
with  
\bee \sum_k C_k^\dag(\e)  \, C_k = (1 + \e^2)I + \e( A_k^\dag B_k + B_k^\dag A_k) \equiv S(\e).
\eee
For sufficiently small $\e$ the operator $S(\e)$ is positive semi-definite and
invertible, and the map $\Phi_{\e}(\rho) =  C_k(\e) S(\e)^{-1/2} \rho   S(\e)^{-1/2} C_k^\dag(\e)$ 
is a CPT map with Kraus operators $C_k(\e) S(\e)^{-1/2}  $.     Thus,
one can find $\e_c$ such that $\e \in (0,\e_c)$ implies that 
 $\Phi_{\e} \in {\cal E}(d_1,d_2)$.   It then follows from
  $\ds{ \lim_{\e \raw 0+} \Phi_{\e} = \Phi}$ that
 $\Phi \in  \ovb{{\cal E}(d_1,d_2)}$.     \qed

When $d_1= 2$, one  can use the singular value decomposition (SVD)
to show that that the Kraus operators of CPT maps with Choi rank at
most two can be written in the form
  \be  A_1 =  \sum_{j = 1,2} \alpha_j |v_j \kb u_j |   \qquad   
   A_2 =  \sum_{j = 1,2}  \sqrt{1 - \alpha_j^2}  |w_j \kb u_j | 
\ee
where $0 \leq \alpha_j \leq 1$, $|u_j \ket$ is pair of orthonormal
vectors in ${\bf C}_2$, and $|v_j \ket, |w_j \ket$ are two pairs of
orthonormal vectors in ${\bf C}_{d_2}$.   This gives all
CPT maps in $\ovb{{\cal E}(2,d_2)}$. 
Although it may seem artificial from a physical point of view to
consider $d_1 \neq d_2$, several reduction results in quantum 
Shannon theory require consideration of maps with $d_1 \neq d_2$.

\begin{prob}
Characterize, classify and/or parameterize the closure
$\ovb{{\cal E}(d_1,d_2)}$ of the set of extreme points of CPT maps
 $\Phi: M_{d_1}  \mapsto M_{d_2}$ for $d_1 > 2$ and $d_2$ arbitrary.
\end{prob}

Although this problem is of some interest in its own right, we will 
give additional motivation in Section~\ref{mult:ext} where we observe
that certain conjectures for CPT maps with $d_1 = d_2 $ can be 
reduced to case of the channels  in the closure of extreme points
with $d_1 \geq d_2$.

\pagebreak 

\section{Convex decompositions of CPT maps or \\
A block matrix generalization of Horn's lemma}

\centerline {\bf Based on joint work with K. Audenaert}  \medskip

Since the set of  CPT map $\Phi: M_{d_1}  \mapsto M_{d_2}$ is convex,
 it can be written
as a convex combination of extreme maps, and one expects that
$d_1^2( d_2^2 -1)$ will suffice.   
For maps on qubits, it was shown in \cite{RSW} that if 
all maps in  $\ovb{{\cal E}(d_1,d_2)}$ are permitted, then only two are
needed and they can be chosen so that the weights are even.
  This result generalizes to any CPT map
with qubit output, i.e., for $\Phi: M_d \mapsto M_2$ one can  write
\be
   \Phi = \half( \Phi_1 + \Phi_2)
 \ee
where $\Phi_1$ and $\Phi_2$ have Choi rank $\leq d$.   We 
conjecture that this result extends to arbitrary CPT maps.

\begin{conj}    \label{conj:AR1}
{\em (Audenaert-Ruskai)} Let $\Phi: M_{d_1}  \mapsto M_{d_2}$
be a CPT map.   One can find $d_2$ CPT maps $\Phi_m$ with Choi rank at most $d_2$
such that
\be    \Phi = \sum_{m=1}^{d_2}  \tfrac{1}{d_2} \Phi_m .
\ee
\end{conj}
The adjoint or dual of a CPT map is a unital CP map and it is useful to
restate the conjecture in this form.
\begin{conj} \label{conj:AR2}
Let $\Phi: M_{d_2}  \mapsto M_{d_1}$
be a CP map with $\Phi(I_2) = I_1$.   One can find $d_2$ unital CP
 maps $\Phi_m$ with Choi rank at most $d_1$
such that
\be   \label{conj:CPunit}
    \Phi = \sum_{m=1}^{d_2}  \tfrac{1}{d_2} \Phi_m .
\ee
\end{conj}
In this form, the conjecture can be viewed as a statement about block matrices, and it
is useful to restate it explicitly in that form.
\begin{conj}  \label{conj:AR3}
Let ${\bf A}$ be a $d_1 d_2$ positive semi-definite matrix
consisting of $d_2 \times d_2$ blocks  $A_{jk}$ each of size $d_1 \times d_1$,
with $\sum_j A_{jj} = M$.   Then one can find $d_2$ block matrices ${\bf B}_m$,
each of rank at most $d_1$, such that $\sum_j B_{jj} = M$, and
\be \label{block1}
  {\bf A} =  \sum_{m=1}^{d_2}  \tfrac{1}{d_2} \, {\bf B}_m
\ee
\end{conj}
If Conjecture~\ref{conj:AR3} holds, then Conjecture~\ref{conj:AR2} 
(and hence Conjecture~\ref{conj:AR1}) 
follows immediately.  One need only let ${\bf A} =  \Phi(\proj{\beta})$ 
be the Choi matrix of $\Phi$ for which $M = \tfrac{1}{d_2} I_{d_2}$.
It would suffice to prove Conjecture~\ref{conj:AR4}  for the case
$M = I_{d_2}$.   The general case then follows by multiplying on the
right and left by the matrix $\tfrac{1}{\sqrt{d_2} }\sqrt{M} \ot I_{d_2}$.
(May be some subtleties if $M$ is non-singular.)

For $d_1 = 1$, Conjecture~\ref{conj:AR3}  is a 
consequence of Horn's Lemma\footnote{See Theorem 4.3.32 of \cite{HJ1}.
Note that \cite{H} is by Alfred Horn, but that \cite{HJ1} is co-authored by Roger A. Horn.}
 \cite{H,HJ1} which says that a necessary
and sufficient condition for the existence of a positive semi-definite
matrix with eigenvalues $\lambda_k$ and diagonal elements
$a_{kk}$ is that $\lambda_k$ majorizes $a_{kk}$.    
\begin{cor}   \label{cor:horn} 
Let $A$ be a $d \times d$ positive semi-definite matrix with ${\rm Tr} \,  A = 1$.
Then there are $d$ normalized vectors ${\bf x}_m$ such that
\be
         A = \sum_{m =1}^{d}  \td {\bf x}_m {\bf x}_m^\dag
\ee
\end{cor} 
\pf  Note that any set of non-negative eigenvalues $\lambda_k$ with
$\sum_k \lambda_k = 1$ majorizes the vector $( \td, \td, \ldots, \td)$.
Therefore, by Horn's lemma, one can find a unitary $U$ and a self-adjoint matrix 
$B$ such that $A = U B^2 U^\dag$ and the diagonal elements
of $B^2$  are all  $\td$.   (In fact, $U, B$ can be chosen to have real elements.)
Write $U = \sum_k {\bf u}_k {\bf e}_k^\dag $
where  ${\bf u}_k$ denotes the $k$-th column of $U$ and ${\bf e}_k$
the standards basis.  Let ${\bf x}_m = \sqrt{d}  \sum_j {\bf u}_j  b_{jm}$.  Then
\be
   A & = & \sum_{jk}  {\bf u}_j  \bra {\bf e}_j B^2 {\bf e}_k \ket {\bf u}_k^\dag  \nn \\
     & = & \sum_{jk} \sum_m  {\bf u}_j  \bra {\bf e}_j B {\bf e}_m \ket
             \bra {\bf e}_m, B  {\bf e}_k  \ket {\bf u}_k^\dag  \nn \\
             & = & \sum_m \td   {\bf x}_m {\bf x}_m^\dag
\ee
and, since the columns of a unitary matrix are orthonormal,   
\be   \label{norm}
     \| {\bf x}_m \|^2  & = &  d \sum_{jk}  {\bf u}_j^\dag \ovb{b}_{jm} b_{km} {\bf u}_k
           = d \sum_{jk} \ovb{b}_{jm} b_{km}  {\bf u}_j^\dag  {\bf u}_k   \nn \\
                & = & \sum_{jk} \delta_{jk}  \ovb{b}_{jm} b_{km} =  d (B^2)_{mm} = d \td = 1.  \qquad \qed
\ee

     This suggests that we restate  the conjecture \eqref{block1} using vectors
     of block matrices of the form
    $   {\bf X}_m^\dag = \pmx  X_{1m }^\dag & X_{2m}^\dag & \ldots & X_{d_2 m}^\dag \emx $
with each block $d_1 \times d_1$. 
 \begin{conj}  \label{conj:AR4}
Let ${\bf A}$ be a $d_1 d_2$ positive semi-definite matrix
consisting of $d_2 \times d_2$ blocks  $A_{jk}$ each of size $d_1 \times d_1$,
with $\sum_j A_{jj} = M$.   Then one can find $d_2$ vectors ${\bf X}_m$ composed
of $d_2$ blocks  $X_{jm}$ of size $d_1 \times d_1$ such that 
\be \label{block2}
  {\bf A} =  \sum_{m=1}^{d_2}  \tfrac{1}{d_2} \, {\bf X}_m {\bf X}_m^\dag, \quad \text{and}
\ee
\be \sum_k  X_{km} X_{km}^\dag= M \qquad  \forall ~ m \ee
\end{conj}
   There is no loss of generality in replacing $B_m$ by ${\bf X}_m  {\bf X}_m^\dag$
with ${\bf X}_m$ of the above form.   If $X$ is $d_1d_2 \times d_1d_2$
with rank $d_1$, then by the SVD it can be written as $X = U D V^\dag$
with $U,V$ unitary and $D$ diagonal with $d_{jj} = 0$ for $ j > d$.   
If $\wtd{D}$ retains only the first $d_1$ columns of $D$, then
   $ \wtd{X}   = U \wtd{D}$ has the desired form 
and $\wtd{X} \wtd{X}^\dag = X X^\dag$.  thus,
Conjecture~\ref{conj:AR4} is clearly a generalization of Horn's lemma to
block matrices.

When $d_2 = 2$, the argument in \cite{RSW}
(due to S. Szarek) is easily extended to give a proof of
Conjecture~\ref{conj:AR3} .
Then  $ {\bf A} > 0$ is equivalent to
\be  \label{rsw1}
   {\bf A} = \pmx \sqrt{ A_{11} }& 0 \\ 0 &  \sqrt{ A_{22}} \emx
   \pmx  I & W \\ W^\dag & I \emx 
   \pmx \sqrt{ A_{11}} & 0 \\ 0 &  \sqrt{ A_{22}} \emx
\ee
with $W$  a contraction.
Write the SVD of W as
\be  \label{rsw2}
    W & = &  U \pmx \cos \theta_1 & 0 &0 &  \ldots&  0 \\
                            0 & \cos \theta_2 & 0 & \ldots & 0 \\
                            \vdots & 0 & \ddots & & \vdots \\
                             0 & 0 & \ldots & 0 & \cos \theta_d \emx V^\dag  \nn  \\
               ~~ \nn \\         & = & 
            \half    U \pmx   e^{i \theta_1} & 0 &0 &  \ldots&  0 \\
                            0 &e^{i \theta_2} & 0 & \ldots & 0 \\
                            \vdots & 0 & \ddots & & \vdots \\
                             0 & 0 & \ldots & 0 &e^{i \theta_d} \emx V^\dag  
                       +      \half      U \pmx  e^{-i \theta_1} & 0 &0 &  \ldots&  0 \\
                            0 &e^{-i \theta_2} & 0 & \ldots & 0 \\
                            \vdots & 0 & \ddots & & \vdots \\
                             0 & 0 & \ldots & 0 &e^{-i \theta_d}  \emx V^\dag \nn \\
               & = & \half (W_1 + W_2)             
\ee
with $W_1$ and $W_2$ unitary.   When $W$ is a $d_1 \times d_1$ unitary, 
$\pmx  I & W \\ W^\dag & I \emx$
has rank $d_1$.  Therefore, substituting \eqref{rsw2} into \eqref{rsw1} shows that
${\bf A} $ is the midpoint of two  matrices with rank at most $d_1$ and
the same blocks on the diagonal as ${\bf A} $.

This argument suggests that one might strengthen the conjecture to require
that each ${\bf B}_m$ have the same diagonal blocks as ${\bf A}$.    However,
this does not appear to hold in the limiting case $d_1 = 1$ with $d_2 > 2$.
In the proof of Corollary~\ref{cor:horn}, it is tempting to replace $B $ by
$C = B V $ with $V$ unitary.   However, in \eqref{norm} we would obtain
$(C^\dag C)_{mm}$ which, unlike $CC^\dag$ need not  have diagonal elements
$\td$.

The original proof of Horn's lemma used a complicated induction argument
based on the properties of augmenting a matrix by a row and column.   Since
we know that \eqref{block2} holds when $d_2 = 2$ or $d_1 = 1$, we have
the starting points for a (probably non-trivial) double induction argument.
Although Audenaert has found 
extensive numerical evidence for the validity of 
Conjectures~\ref{conj:AR1}-\ref{conj:AR4},
a proof seems to be elusive.


\section{Generalized depolarized channels}

\subsection{Depolarized Werner-Holevo channels}  \label{sect:WH}

The Werner-Holevo channel    ${\cal W}(\rho) = \tfrac{1}{d-1}\big( (\tr \rho) \,   I - \rho^T \big)$
has been extensively studied, especially in connection with the conjectured
mutliplicativity of the maximal output $p$-norm, defined as 
$\nu_p(\Phi) = \sup_{\rho} \| \Phi(\rho) \|_p$.
For $d = 3$, the maximal output $p$-norm is not multiplicative for $p > 4.79$.
However, it is known that $\nu_p({\cal W} \ot ({\cal W}) = [ \nu_p({\cal W})]^2$
for $1 \leq p \leq 2$.    For larger $d$ one obtains a counter-example to
multiplicativity only for correspondingly large $p$.   In fact, it has been 
argued \cite{GLR}   that for $d > 2^p$ the WH channel is multiplicative.

${\cal W}$ maps any pure state $\proj{\psi}$   
to $\tfrac{1}{d-1} E$ with $E = I - \proj{\psi}$.   Therefore, 
when $d$ is large, ${\cal W}$ behaves much like the completely
noisy map (although it is never EB).   It is natural to consider channels
of the form
\be  \label{dWH}
    \Phi_x =  x \id + (1-x) {\cal W}
\ee
and ask if they also satisfy the multiplicativity conjecture \eqref{mult} for $1 \leq p \leq 2$.
Channels of the form \eqref{dWH} were considered by Ritter \cite{Ritt} in
a different context.
\begin{prob}  \label{prob:WHdep}
   Show that the channel $ \Phi_x =  x \id + (1-x) {\cal W}$ satisfies the 
   multiplicativity property  $\nu_p(\Phi_x) \ot (\Phi_x) = [ \nu_p(\Phi_x)]^2$
   for $1 \leq p \leq 2$.
\end{prob}

When $d = 3$ and $x = \tfrac{1}{3}$, the channel \eqref{dWH} becomes
\be  \label{WH13}
     \Phi_{1/3}(\rho) = \tfrac{1}{3} \big( I + \rho - \rho^T )
\ee
which has many interesting properties.   It seems to have been
first considered  by Fuchs, Shor and Smolin, who  published only
 an oblique remark at the end of \cite{Fuchs}.    They wrote it
in a very different form, which is also given in \cite{HSR}.  
 Let $|1 \ket, |2 \ket,  |3 \ket$ be an orthonormal basis for ${\bf C}_3$
 and define
 \bee
     |\psi_0 \ket & = &  3^{-1/2} \big( |1 \ket   + |2 \ket +  |3 \ket \big) \\
      |\psi_1 \ket & = &  3^{-1/2} \big( |1 \ket   - |2 \ket -  |3 \ket \big) \\
       |\psi_2 \ket & = &  3^{-1/2} \big( |1 \ket  - |2 \ket +  |3 \ket \big)  \\
        |\psi_3 \ket & = &  3^{-1/2} \big( |1 \ket  + |2 \ket -  |3 \ket \big) .
 \eee
Now let $\Psi$ be the channel whose Kraus operators are
$\tfrac{\sqrt{3}}{2}  \proj{\psi_k} $ for $k = 0,1,2,3$.   
This channel has the following properties:
\begin{enumerate}

\item   $\Psi = \Phi_{1/3} = \tfrac{1}{3} \id + \tfrac{2}{3} {\cal W}$.
  Although this is not obvious, it is easily verified and implies
   \eqref{WH13}.
Thus, $\Psi$ maps every real density matrix to the maximally
mixed state.   

\item  $\Psi$ is unital and the Holevo  capacity satisfies
\be
    C_{\rm Hv}(\Phi) = \log 3 - S_{\min}(\Phi)
\ee
but requires $6$ (non-orthogonal) input states to achieve this capacity.  
It is not hard to see that $S_{\min}(\Phi)$ is achieved on inputs which
are permutations of  $(1, \pm i, 0)^T$.

\item  $\Psi$ is an extreme point of the EB channels which is
neither CQ nor an extreme point of the CPT maps \cite{HSR}.

\end{enumerate}

A solution of Problem~\ref{prob:WHdep} in the case $p = 2$
was recently reported by Michalakis \cite{M}.

\subsection{Further generalizations of depolarization}

In \cite{Wp} channels  ${\cal M}_{\epsilon} $ which whose output is 
always close to a maximally mixed
state in the sense  $\| \Phi(\rho) - \td I || < \epsilon$ play an important role.
It seems natural to define a polarization of such channels
\be
      \Phi_{x,\e} = x \id + (1-x) {\cal M}_{\epsilon} 
\ee
For $x$ close to 1, one explects multiplicativity to holds, and it is 
natural to ask several questions.

\begin{prob}
Does $ \Phi_{x,\e} $ satisy \eqref{mult} for $1 \leq p \leq 2$?  For sufficiently small $\e$?
\newline  \medskip
If not,  for what values of $x$ and/or  $p$ does \eqref{mult} holds and
how do they depend on $\epsilon$?  
\end{prob}

\section{Random sub-unitary channels}

We now introduce a class of extreme points motivated by the WH channel.

The Kraus operators for the WH channels with $d = 3$ can be written as
\be
   A_k =   \half   X^k \pmx 0 & 1 & 0 \\ -1 & 0 & 0 \\ 0 & 0 & 0 \emx  \qquad  k = 0,1,2
\ee
where $X$ is the shift operator $X |e_j \ket = |e_{j+1} \ket$.    This suggests
a natural generalization to channels with Kraus operators
\be    \label{Kr2}
A_k =  \half   X^k \pmx u_{11} & u_{12} & 0 \\ u_{21} & u_{22} & 0 \\ 0 & 0 & 0 \emx =
    \half   X^k \pmx U & 0 \\ 0 & 0 \emx   \qquad  k = 0,1,2
\ee
with $u_{jk}$ the elements of a $2 \times 2 $ unitary $U$.   The choice 
$U = \pmx 0 & 1 \\ 1 & 0 \emx$ does not give a counterexample to \eqref{mult},
 although the
effect of a tensor product on a maximally entangled state is the same as
the WH channel.   This is because changing $-1$ to $+1$ allows a ``purer''
 optimal output for a single use of the channel; to be precise, for $ +1$ the input 
 $\tfrac{1}{\sqrt{3}} (1, 1, 1)$
 yields an output with eigenvalues $\tfrac{2}{3}, \tfrac{1}{6}, \tfrac{1}{6}$
 as compared to eigenvalues $\half, \half, 0$ for $-1$.

By contrast, the standard generalization of the WH channel to $d > 3$
involves $\binom{d}{2}$ choices of $ \pmx 0 & 1 \\ -1 & 0 \emx$ as the
only non-zero block of a $d \times d$ matrix.   It would seem natural to
study channels with $d$ Kraus operators of the form
\be    \label{Kr3}
     \tfrac{1}{d-1} X^k \pmx U & 0 \\ 0 & 0 \emx   \qquad  k = 0,1, \ldots d \mm 1.
\ee
where $U$ is a  $ d \mm 1 \times d \mm 1 $ unitary matrix.
Such channels are  generically extreme and always in the
closure $\ovb{{\cal E}(d,d)}$.
Limited attempts to find new counter-examples of this type have found
similar behavior to changing $+1$ to $-1$; they have
outputs which are ``too pure'' for a single use of the channel.

Nevertheless, channels with Kraus operators of the form \eqref{Kr3}
have interesting properties that makes them worth further study.
Moreover, it is not necessary to use the same $U$ in every Kraus
operator.  One can choose
\be  \label{subu}
    A_k = \tfrac{1}{d-1} X^k  \pmx U_k & 0 \\ 0 & 0 \emx   \qquad  k = 0,1, \ldots d \mm 1.
\ee
with $U_k$ any set of unitaries in $M_{d-1}$.
     With a few exceptions, channels
whose Kraus operators have the form \eqref{subu} are extreme points of the
CPT maps on $M_d$, and are  always in ${\ovb{\cal E}(d,d)}$. 

The WH channel gives a counter-example to multiplicativity for large
$p$ because maximally entangled states have outputs
whose $p$-norms are relative maxima of  \linebreak $\| ({\cal W} \ot {\cal W} )(\rho) \|_p$,
Nathanson \cite{NR} has shown analytically that for any $p$ the output of
any maximally entangled state gives a critical point, but
Shor has found numerical evidence \cite{ShorPC} that this is a relative maximum
only for $p \geq 3$.    This suggests that
one look at  other random sub-unitary channels.
\begin{prob}
Let $\Phi$ be a channel with Kraus operators of the form \eqref{subu}.
Does the set of  relative maxima of  $\| (\Phi \ot \Phi)(\rho) \|_p$ always include
outputs whose input is maximally entangled?   If not, for what $p$ and under
what circumstances do maximally entangled inputs yield outputs which are
relative maxima?
\end{prob}
Despite the failure of Ruskai's very limited attempt to find
new counter-examples of this type for $d = 4,5$, more extensive numerical
investigations, perhaps with different, randomly chosen, $U_k$,  could be worthwhile.   
Further suggestions about numerical searches are given in Section~\ref{sect:newcx}.
Even a negative result could provide some insight.  
\begin{prob}
Search for new counterexamples to \eqref{mult} with
$\Phi$ a channel with Kraus operators of the form \eqref{subu}.
\end{prob}

In addition to looking at the optimal output purity of these channels,
one can also ask about their coherent information and quantum capacity.
\begin{prob}
What are the properties of the coherent information of random sub-unitary channels?
When are they degradable?    When is their coherent information additive?
\end{prob}

\noindent{\bf Remark:} (added 11 August 2007).   There has been recent interest
in the question of multiplicativity of minimalÊ output rank \cite{HLW,CMW}.   For
$d = 4$, the sub-unitary channel  $\Phi$ with $3 \times 3$ unitary operators 
corresponding to the permutations
$(123),  (134),  (142),  (243)$ has minimal output rank 3.   The channel  
$\Phi \ot \Phi$ acting on a maximally entangled state has output rank 10,
which does not give a violation.    However, the behavior of this channel
suggests that numerical investigations of similar examples for somewhat
higher $d$ might be worth investigating for counter-examples to the
multplicativity question for $p < 1$ as discussed further in Section~\ref{sect:newcex}.

\section{Additivity and multiplicativity conjectures}

\subsection{Prelude}

Soon after the 14 June 2007 version of this manuscript was posted on the 
BIRS web site, counter-examples were found to the multiplicativity conjectures
for all $p > 1$ \cite{Wp,Hp}.     Nevertheless, the additivity conjectures and
many related questions remain open.  Therefore, I have made only minor
changes to most of this section and discuss the recent developments and
new questions they raise in Section~\ref{sect:newcex}.     Moreover, these
 existence of counter-examples also raises new questions.
Thus,   there may still be value in some of the old material,  such as Theorem~\ref{thm:opt2}.

\subsection{The conjectures}

One of the outstanding open problems in quantum information is the
 additivity of minimal output entropy, i.e.,
 \be   \label{Smin}
  S_{\min}(\Phi \ot \Omega) =   S_{\min}(\Phi ) +  S_{\min}(\Omega)
 \ee
where  $ \ds{S_{\min}(\Phi) = \inf_\gamma S[\Phi(\gamma)}$ where
the infimum is taken over the set of density matrices $\gamma$
so that $\gamma > 0$ and $\tr \gamma = 1$.    This conjecture has
considerable importance because Shor \cite{ShorEq} has shown that
it is globablly equivalent to the conjectured additivity of Holevo capacity
and several conjectures about entanglement of formation.
Shirokov \cite{Shir1,Shir2} has even shown that additivity in all
finite dimensions would have implications for certain infinite dimensional
channels.   Fukuda \cite{F} and Wolf \cite{FW} have given some
additional reductions.
 
Amosov, Holevo and Werner \cite{AHW} realized that \eqref{Smin} would
follow if the following conjecture holds for $p \in (1,1+\epsilon)$ with
$\epsilon > 0$.
 \be  \label{mult}
     \nu_p(\Phi \ot \Omega) =  \nu_p(\Phi )  \nu_p(\Omega)
 \ee
where $\ds{ \nu_p(\Phi) = \inf_\gamma \| \Phi(\gamma) \|_p}$.   Although,
Werner and Holevo \cite{WH} found a counter-example to \eqref{mult} for
large $p$, it seems reasonable to conjecture that \eqref{mult} holds
for $1 \leq p \leq 2$.     This conjecture can be rewritten  \cite{AF,Hp} using the Renyi
entropy, which is essentially the difference quotient at $p = 1$ of  $p \log  \| \gamma \|_p$,
i.e.,
\be   \label{renyi}
    S^p(\gamma) \equiv  \tfrac{1}{p-1} \log \tr \gamma^p.
\ee
This expression is meaningful  for any $p \geq 0$ with the
understanding that   $S^0(\gamma) = \log \hbox{rank}(\gamma)$ and
$S^1(\gamma)$ the usual von Neumann entropy.  
Then \eqref{mult} can be rewritten as
\be  \label{padd}
    S^p_{\min}(\Phi \ot \Omega) =   S^p_{\min}(\Phi ) +  S^p_{\min}(\Omega)
\ee
with $\ds{S^p_{\min}(\Phi ) = \inf_{\gamma}   S^p[\Phi(\gamma)]}$.

\subsection{Finding counter-examples}   \label{sect:newcx}

 It is surprising that no counter-example  to  \eqref{mult}
is known other than the WH channel \cite{WH} and very small perturbations of it.
Moreover, one has no counter-example for $p < 4.79$.    
Some authors \cite{KR} have conjectured that  \eqref{mult} holds for $1 \leq p \leq 2$. 
If so, one would expect to have a family of counter-examples for $p > 2$.    
More generally, if the conjecture holds for $1 < p < p_c$, one would expect 
to find counter-examples for $p > p_c$  arbitrarily close to to $p_c$. 
\begin{prob}    \label{p2cex}
Find more counter-examples to \eqref{mult}.   Do they suggest that
the conjecture holds for $1 \leq p \leq 2$?
\end{prob}

One strategy for finding new counter-examples, is to first search numerically
for additional counter-examples for very {\em large} $p$ using Theorem~\ref{thm:opt2}
below.     For any new examples found, study the critical points numerically
and determine the values of $p$ for which one ceases to have a counter-example
and for which one ceases to even have a relative maximum for entangled inputs.
Perhaps this will give some insight into the nature of  counter-examples
that will allow one to find some in the range $2 < p < 4.79$.    The reason for starting with
large $p$ is that the algorithm for finding relative maxima using Theorem~\ref{thm:opt2}
is faster and more robust for large $p$.

The following extension of  Shor's algorithm for finding relative minima of the
minimal ouput entropy (see Appendix~  of \cite{DR})   was proved by C. King
using  H\"older's inequality in the case $p > 1$.    We present a different proof,
valid for all $p > 0$.
We first note that Shor's argument uses the positivity of relative entropy,
which is based on Klein's inequality, using the more general form in 
Ruelle \cite{R} for convex functions  
\be
    \tr f(A) - \tr f(B) \geq  \tr (A-B) f^\prime(B)
\ee
where $A,B$ are positive semi-definite matrices.   Since the function $f(x) = x^p$
with $p > 1$ is convex, this gives
\be   \label{Kp}
    \tr A^p - \tr B^p  \geq  p(A-B) B^{p-1} .
\ee
 \begin{thm}  \label{thm:opt2}
Let $p > 0$ and $\Omega$  a CPT map with $\wh{\Omega} $ its adjoint with respect to the
  Hilbert-Schmidt inner product.   For fixed $\gamma = \proj{\psi_0}$ and
   $\Big  \{   \begin{array}{c} p < 1 \\ p > 1  \end{array} \Big\}$, let $\psi_1$ 
    be the eigenvector corresponding
  to the   $\Big  \{   \begin{array}{c}  \hbox{smallest} \\  \hbox{largest}  \end{array} \Big\}$  
  eigenvalue of $\wh{\Omega}\Big[  \Omega(\gamma) \Big]^{p-1}$.
  Then   $\norm{ \Omega( \proj{ \psi_1 })}_p  ~  \Big  \{  \begin{array}{c}  \leq  \\  \geq
    \end{array} \Big\}  ~  \norm{ \Omega(\proj{ \psi_0)}}_p$
 \end{thm}

\pf   First consider $p > 1$.   The max min principle implies that
\be 
\bra \psi_1  \wh{\Omega}\big[  \Omega\big( \proj{ \psi_0}\big) \big]^{p-1} \psi   \ket \geq
         \bra \psi_0  \wh{\Omega}\big[  \Omega \big(\proj{ \psi_0}\big) \big]^{p-1} \psi_0  \ket.
         \ee
         which can be rewritten as
         \be  \label{mm} 
           \tr  \Omega\big( \proj{ \psi_0}\big) \big[\Omega\big( \proj{ \psi_0}\big)]^{p-1}  \geq
              \tr  \big[\Omega\big( \proj{ \psi_0}\big)]^p
         \ee
 Then using \eqref{Kp} with   $A = \Omega\big( \proj{ \psi_1}\big),  B = 
   \Omega\big( \proj{ \psi_0}\big) $  gives
\bee
  \lefteqn{    \tr    \big[\Omega\big( \proj{ \psi_1}\big)]^p -  \tr  \big[\Omega\big( \proj{ \psi_0}\big)]^p } 
   \qquad \\
  &    \geq &  p  \Big( \tr  \Omega\big( \proj{ \psi_0}\big) \big[\Omega\big( \proj{ \psi_0}\big)]^{p-1} -
              \tr  \big[\Omega\big( \proj{ \psi_0}\big)]^p \Big)  ~ \geq  ~ 0.
\eee
where the last inequality follows from \eqref{mm}.   Taking  $p$-th roots gives
the desired result.  

For $0 < p < 1$, the function $f(x) = x^p$ is concave and the same argument
goes through with all inequalities reversed.  \qquad
      \qed

Using this result repeatedly with $\psi_{k+1}$   the eigenvector corresponding
  to the largest eigenvalue of $\wh{\Omega}\big[  \Omega(\proj{\psi_k}) \big]^{p-1}$,
  gives a  sequence converging to a relative maximum of $\norm{ \Phi(\gamma) }_p$.

\subsection{Specific multiplicativity problems}

Proving multiplicativity of the depolarized WH channel was already
mentioned in Section~\ref{sect:WH}.   Recently,
Michalakis reported \cite{M} a proof for $p=2$.   In view of 
the fact that some depolarized WH channels do {\em not} satisfy the
very unappealing conditions based on positive entries  used in \cite{KR,KNR},
the approach in \cite{M} may be useful in investigating other classes of channels.
\begin{prob}
For what classes of channels can \eqref{mult} be proved for $p = 2$.
\end{prob}  

In \cite{NR}, a class of channels is defined using mutually unbiased
bases, with each basis defining an``axis''.    These channels can
be described by ``multipliers'' in a manner similar to unital qubits
channels, and when all multipliers are non-negative they seem
very similar.    However, even for a single use of  a channel 
 some questions are open.   See Conjecture 9 of \cite{NR}.
 If this conjecture is true, then additivity and multiplicativity can
 be reduced to the case of  ``maximally squashed'' channels
 which are generalizations of the two-Pauli qubit channel.
\begin{prob}
Find a proof of multiplicativity for the two-Pauli qubit channel,
which does not use unitary equivalence to channels with negative 
multipliers.
\end{prob}   

Since most recent investigations of additivity \eqref{Smin} 
have approached the problem through the multiplicativity conjecture,
it is worth noting that Amosov has obtained some results \cite{Am1,Am2}
in special cases by a very different approach using the montonicity of 
relative entropy.     Also recall that Shor's proof \cite{ShorEB}
of additivity for entanglement breaking channels used entropy arguments
based on strong subadditivity.    Although these basic properties of entropy
are unlikely to suffice for more general channels, they do demonstrate that
multiplicativity is not the only route to additivity.

\subsection{Reduction to extreme points}  \label{mult:ext}

Although the set of CPT maps $\Phi:M_{d_A} \mapsto M_{d_B}$
is convex, one can not use convexity to reduce additivity or 
multiplicativity to that of the extreme channels.   One can, however,
use the notion of complementary channels to obtain a kind of
global reduction to extreme channels.

The notion of complementary channel was first used in quantum
information theory in a paper of Devetak and Shor \cite{DS} and then
studied in detail in \cite{Hv,KMNR}.    This concept is equivalent
to one obtained much earlier in a more general context by
Arveson \cite{Arv} in the section on lifting commutants.  
(See the appendxi to \cite{CRS} for details.)

   If $\Phi:M_{d_A} \mapsto M_{d_B}$, its
complement is a CPT map $\Phi^C:M_{d_A} \mapsto M_{d_E}$ with
Choi rank $d_B$.   Whenever $d_B \leq d_A$, the complement belongs
to the class of generalized extreme points.    Therefore,   the results
in \cite{Hv,KMNR} imply that if we can prove additivity  for all
maps in $\ovb{{\cal E}(d_1,d_2)}$, it will hold for all CPT maps with 
$d_B \leq d_A$.     Moreover, Shor's channel extensions \cite{ShorEq}
used to establish the equivalence of various additivity results 
increase only $d_A$.   Hence, additivity for tensor products of
all extreme maps with $d_A \geq d_B$  would imply it for all maps
with $d_A = d_B$.    

\begin{prob}
Identify new classes of extreme CPT maps for which additivity and/or 
multiplicativity can be proved.
\end{prob}

\begin{prob}
Can one prove \eqref{mult} for random sub-unitary channels, at least for $p = 2$.
If not, do these channels provide additional counter-example?
\end{prob}

\subsection{New counter-examples and their implications}  \label{sect:newcex}

Very recently (July, 2007), Winter \cite{Wp} solved Problem~\ref{p2cex} by showing
the existence of  counter-examples for all $p > 2$.   Moreover,
his approach failed at $p = 2$, which seemed to provide support for the validity of 
multiplicativity in the range $1 < p  \leq 2$.    But soon after, 
Hayden \cite{Hp} showed that there exist counter-examples for 
$1 < p < 2$ and this was extended to  $p = 2$ by Winter.

Hayden \cite{Hp} also provided an analysis of his examples that indicates
 that \eqref{Smin} still holds for these channels and suggested that one
 try to establish additivity by proving \eqref{padd} for $p < 1$.   King \cite{Knew}
 announced that his arguments for multplicativity of entanglement breaking channels
 \cite{King4} extend to $0 < p < 1$.    He also observed that the proofs
 of \eqref{mult} for   unitalÊ qubit channels 
 \cite{King2} and depolarizing channels, \cite{King3}
were basd on the following  inequality of Lieb-Thirring \cite{LT}   
 \be
       \tr (AB)^p \leq \tr A^p B^p
 \ee
 for $ p \geq 1$ and $A,B$ positive semi-definite.  Since Araki \cite{Ak} has shown that the reverse
 inequality is valid for $ 0 \leq p < 1$, his results for unital qubit and
 for general depolarizing channels should also readily extend to $0 < p < 1  $.
 
 However, hopes for validity of \eqref{padd} for $0 \leq p < 1$ were
 shattered when Harrow, Leung and Winter \cite{HLW} announced 
 counter-examples for $p = 0$.    These examples differ from those for $p > 1$.
But they are also based on the results and methods introduced in \cite{HLW}  
 on the prevalence of nearly maximally entangled states in large dimension.
 It seems only a matter of time until counter-examples are shown to exists
 for any $p \in (0,1)$.   
 
 Nevertheless, it is worth emphasizing that none of the counter-examples
 are uniform in $p$, i.e., as $p$ approaches $1$ the counter-example fails
 and a new one must be found with dimension increasing to infinity as $p \mapsto 1$.
 Thus, the following much weaker forms of   \eqref{mult} and \eqref{padd} are
 not excluded.     The validity of any one of the four conjectures which  
 follow would imply that \eqref{Smin}
 and all the equivalent additivity conjectures hold.  
  \begin{conj}
 For any fixed pair of channels $\Phi, \Omega$, there is a $p^* > 0$ such that either
 
 (i)  $p^* < 1$ and \eqref{padd} holds for all $p \in (p^*,1)$, or
 
 (ii)   $p^* > 1$ and \eqref{padd} holds for all $p \in (1, p^*)$. 
 \end{conj}
 
  \begin{conj}
 For any fixed  integer $d$ there is  a $p_d > 0$ such that either
 
 (i)  $p_d < 1$ and \eqref{padd} holds whenever $\Phi: M_{d^\prime} \mapsto M_{d^\prime}$, 
  ${d^\prime} \leq d$ and $p_d < p < 1$, or
 
 (ii)   $p_d > 1$ and \eqref{padd} holds whenever $\Phi: M_{d^\prime} \mapsto M_{d^\prime}$, 
  ${d^\prime} \leq d$ and $1 < p  < p_d $
  \end{conj}
For brevity  we stated  the  conjectures above
 in pairs, but in each case the form (i) or (ii) is a separate conjecture. 
  
  If the additivity conjectures are true, proving either of the above conjectures 
  seems less likely than proving \eqref{Smin} directly.    Moreover, Shor's
  channel extension methods give global equivalences which require 
  consideration of 
  CPT maps $\Phi :M_d \mapsto M_{d^\prime}$ with $d \geq d^\prime$.
  Thus one should extend the above conjectures to include the case 
  $d > d^\prime$.   However, we preferred to state them in the simpler form.

Although based on similar techniques, the actual form of the channels giving
counter-examples for $p > 1$ and $p < 1$ seems to be different.
This leads to 
 \begin{prob}
    Does there exists a channel or pair of channels that violates \eqref{padd}
    for both some $p_1 > 1$ and some $p_2$ with $0 < p_2 < 1$?
 \end{prob}
If the answer is negative,  then \eqref{Smin} holds because one can always  
approach $p = 1$ from either above or below.    This seems a rather unlikely
approach to proving  \eqref{Smin}, but thinking about it might  provde
   some insight   about this additivity conjecture.

The need for large dimensions to find counter-examples raises the
 question of whether channels for smaller dimensions might
satisfy \eqref{padd} for two copies, but not for a large number.
\begin{prob}
Find an example of a channel $\Phi$, an  integer $m$ and a $p > 0$ such that 
   $S_{\min}^p(\Phi^{\ot n} ) = n S_{\min}^p(\Phi)$ for $n < m$ but
   $S_{\min}^p(\Phi^{\ot m} ) < m  S_{\min}^p(\Phi)$.
   \end{prob} 
Current results do not even exclude the possibility that
a non-unital qubit channel violates additivity for $\Phi^{\ot m}$.    
Curiously, \eqref{mult} has only
been proved \cite{KK} for non-unital qubit maps when $p = 2$ or $p \geq 4$.

All of the counter-example  results obtained thus far are given as existence
theorems.   It would be useful to have explicit counter-examples.
\begin{prob}
Find explicit examples of channels which violate \eqref{mult} for $p \neq 1$.
\end{prob}

In the case of Winter's examples \cite{Wp} for $p > 2$, one can show that
the so-called CB entropy \cite{DJKR} is positive and the coherent 
information is negative.   (When the coherent information is achieved with
a maximally entangled state, the CB entropy and coherent information
differ only by a sign change.)  Thus, although these channels are not
entanglement  breaking (EB), they preserve  very little entanglement -- not even
enough to allow one to recover a single EPR pair in the sense of 
Horodecki, Oppenheim and Winter \cite{HOW1,HOW2}.   The WH counter-examples
also have positive CB entropy except for $d = 3$ when it is exactly zero.
Thus, for $p > 2$, the known counter-examples suggest that 
entanglement does not enhance the optimal output purity until the
channel is very close to EB.    One can ask if this holds for other examples,
particularly those for $p < 2$.

\begin{prob}
Do all counter-examples to multiplicativity \eqref{mult} have  non-negative 
CB entropy and/or zero coherent information?
\end{prob}

Finally, one can ask whether or not additivity itself holds.   It is worth
recalling that  the equivalent capacity conjecture was stated in \cite{BFS}
in a form that seemed to favor superadditivity.    Thus, the  ultimate open
question is still.
\begin{prob}
  Prove \eqref{Smin} or find a counter-example.
\end{prob}

\section{Coherent information and degradability}

In  \cite{CRS} on degradability several questions were raised
of which we mention one.
\begin{prob}
Find pairs of channels  ${\cal M, N}$ that are mutually degradable in
the sense that there exist channels ${\cal X, Y}$ such that
\be
    {\cal X} \circ {\cal M} = {\cal N}^C  \qquad    {\cal Y} \circ {\cal N} = {\cal M}^C.
\ee
\end{prob}
At present, the only examples known have ${\cal M} = {\cal I}$ which is
universal in the sense that ${\cal N}$ is arbitrary.   This works because
${\cal I}$ is universally degradable and its complement $\tr $ is a universal
degrador.   Can other examples be found?    It may be that when  ${\cal N}$ 
has Choi rank $d^2$, one must have ${\cal M} = {\cal I}$.   Therefore, it 
seems worth looking for examples in which both have lower Choi-rank.
It would be particularly interesting to find pairs in which both have 
Choi-rank $d$, but are not individually degradable.

\section{Local invariants for $N$-representability}

In the 1960's a variant of the quantum marginal problem known as
$N$-representability attracted considerable interest.   The question 
is to find necessary and sufficient conditions on a $p$-particle
reduced density matrix  $\rho_{1,2, \ldots t}$ in order that there exists an
 anti-symmetric (or symmetric for bosons) 
 $N$-particle density matrix $\rho = \rho_{1,2 \ldots N}$  such that 
$\trp_{t+1,t+2, \ldots N}  \rho_{1,2 \ldots N} = \rho_{1,2, \ldots t}$.
The pure $N$-representability problem, for which one requires that 
the preimage  $ \rho_{1,2 \ldots N} = \proj{\Psi}$
come from an anti-symmetric (or symmetric) pure state $| \Psi \ket$ is also of interest.

A full solution was found only to the mixed state problem for the one-particle 
density matrix, for which it is necessary and sufficient that the eigenvalues
of $\rho_1$ are $\leq \tfrac{1}{N}$ when $\tr \rho_1 = 1$.       Other results were
obtained for a few very special situations, and some reformulations were
found.   For the two-particle reduced density matrix, a collection of necessary inequalities
were obtained, but little else was known.   For over 30 years, there was very
little progress until two recent breakthroughs.    Klyachko \cite{K1} solved
the pure state $1$-representability problems.   Liu, Christandl and Verstraete 
\cite{LCV}
showed that some version are QMA complete. 

Although many open questions remain, we consider only one which
may be amenable to quantum information theorists.    As Coleman
pointed out, $N$-representability must be independent of the $1$-particle
basis used to write the density matrix, i.e., the solution can be expressed
in terms of what one might call local invariants.    These are parameters
which are invariant under transformations of the form 
$U \ot U \ot \ldots \ot U = U^{\ot p}$.     For the $1$-matrix, these are just unitary
invariants, which are known to be the eigenvalues.     For $p = 2$ the set of
local invariants includes the eigenvalues, but must contain other parameters as
well.   Surprisingly, no complete set of local invariants in which $N$-representability
conditions for the $2$-matrix can be expressed is known.
\begin{prob}
Find a minimalÊ complete set of local invariants for an anti-symmetric
(or symmetric) $2$-particle density matrix.
\end{prob} 
 


\begin{thebibliography}{~~}

 \bibitem{AF}R. Alicki and M. Fannes, 
``Note on multiple additivity of minimal entropy output of extreme 
$SU(d)$-covariant channels''
{\em Open Systems and Information Dynamics} {\bf 11}, 339--342 (2004).
quant-ph/0407033.

              
\bibitem{Am1} G.G. Amosov,
``On Weyl channels being covariant with respect to the maximum commutative group of unitaries''
{\em J. Math. Phys.}  {\bf  48},   2104--2117 (2007).   arXiv:quant-ph/0605177 


\bibitem{Am2} G.G. Amosov,
``The strong superadditivity conjecture holds for the quantum depolarizing channel in any dimension''
{\em  Phys. Rev. A}  {\bf 75},   060304 (2007)   arXiv:0707.1097

\bibitem{AHW}  G.~G. Amosov, A.~S. Holevo, and R.~F. Werner, 
``On Some Additivity Problems in Quantum Information Theory'', 
{\em Problems in Information Transmission}, 
{\bf 36}, 305--313 (2000).  eprint  math-ph/0003002

\bibitem{Ak} H. Araki,  ``On an inequality of Lieb and Thirring''
{\em Lett. in Math. Phys.}  {\bf 19}, 167--170 (1990).

\bibitem{Arv} W. Arveson,
``Subalgebras of C*-Algebras''
  {\em  Acta Math.}   {\bf 123},  141--224 (1969).
  
  \bibitem{BFS} C. H. Bennett, C. A. Fuchs and J. A. Smolin,
``Entanglement-enhanced classical communication on a noisy quantum
channel'', 
{\em Quantum Communication, Computing and Measurement},
edited by O.~Hirota, A.~S. Holevo, and C.~M. Caves (Plenum Press,
NY, 1997), pages~79--88.
(quant-ph/9611006)


\bibitem{Choi} M-D Choi,
``Completely Positive Linear Maps on Complex Matrices''
 {\em Lin. Alg. Appl.} {\bf 10}, 285--290 (1975).
 
  \bibitem{Cole} A.J. Coleman,
 ``The structure of fermion density matrices''
 {\em Rev. Mod. Phys.} {\bf 35}, 668--687 (1958).
 
\bibitem{CMW}  T. S. Cubitt, A. Montanaro and  A. Winter, 
``On the dimension of subspaces with bounded Schmidt rank''
arXiv:0706.0705
 
 \bibitem{CRS} T. Cubbit, M.B. Ruskai and G. Smith,
 in preparation
 
\bibitem{DR} N. Datta and M.B. Ruskai,
 ``Maximal output purity and capacity for asymmetric unital qudit channels''
 {\em J. Phys. A: Math. Gen.}  {\bf  38},  9785--9802 (2005). (quant-ph/0505048)

\bibitem{DFH} N. Datta, M. Fukuda and A.S. Holevo,
                   ``Complementarity and additivity for covariant channels"
                   {\it Quant. Info. Proc.}  {\bf 5},  179--207 (2006).
                   
                   \bibitem{DJKR}  I. Devetak, M. Junge, C. King,  and M.~B. Ruskai,
 ``Multiplicativity of completely bounded p-norms implies a new additivity result''
 {\em Commun. Math. Phys.}   {\bf 266}, 37--63 (2006).   (quant-ph/0506196).

 \bibitem{DS}  I. Devetak and  P. W. Shor
``The capacity of a quantum channel for simultaneous transmission of classical and quantum information''  
{\em Commun. Math. Phys.}  {\bf 256},  287--303   (2005).
 quant-ph/0311131
 
\bibitem{Fuchs} C. Fuchs, ``Nonorthogonal quantum states maximize classical
information capacity", {\em Phys. Rev. Lett. }  
{\bf 79}, 1162--1165 (1997).

\bibitem{F}   M. Fukuda,
``Simplification of additivity conjecture in quantum information theory''
  arXiv:quant-ph/0608010

\bibitem{FW}   M. Fukuda and M.M. Wolf,
``Simplifying additivity problems using direct sum constructions'' 
arXiv:0704.1092

\bibitem{GLR} V. Giovannetti, S. Lloyd and M. B. Ruskai,
``Conditions for multiplicativity of maximal $l_p$-norms of channels
for fixed integer $p$'', {\em J. Math. Phys.}  {\bf  46}, 042105 (2005).
quant-ph/0408103.

\bibitem{HLW}  A. Harrow, D. Leung and A. Winter,
private communication

\bibitem{Hp} Patrick Hayden 
``The maximal p-norm multiplicativity conjecture is false''
arXiv:0707.3291     

\bibitem{HLW} P. Hayden. D. Leung and A. Winter,
 ``Aspects of generic entanglement''
 {\em Commun. Math. Phys.}  {\bf 265}, 95--117 (2007).


\bibitem{H} A. Horn, 
``Doubly stochastic matrices and the diagonal of a rotation matrix''
{\em  Amer. J. Math. }  {\bf 76},  620--630(1954).  

 \bibitem{HJ1} R.A. Horn and C.R. Johnson, {\em Matrix Analysis}
(Cambridge University Press, 1985).

\bibitem{Hv} A. S. Holevo, 
  ``On complementary channels and the additivity problem''
  quant-ph/0509101 published as part of \cite{DFH}.
  
\bibitem{HOW1}  M.ÊHorodecki, J. Oppenheim and Andreas Winter
``Partial quantum information'' {\em Nature},  {\bf 436}, 673--676 (2005);
posted as ``Quantum information can be negative''  quant-ph/0505062

\bibitem{HOW1}  M.ÊHorodecki, J. Oppenheim and Andreas Winter,
``Quantum state merging and negative information''
{\em Commun. Math. Phys.}  {\bf 269},  107--136   (2007).


\bibitem{HSR}	M.ÊHorodecki,  P. Shor, and M. B. Ruskai 
 ``Entanglement Breaking Channels''  
  {\em Rev. Math. Phys }  {\bf 15},  629--641 (2003).  
	  (quant-ph/030203)
	  

\bibitem{King2} C. King,
``Additivity for unital qubit channels'',
{\em J. Math. Phys.} {\bf 43}, no. 10 4641--4653  (2002).

\bibitem{King3} C. King, ``The capacity of the quantum depolarizing channel'',
{\em IEEE Trans. Inform. Theory}  {\bf 49}, no. 1 221--229, (2003).

\bibitem{King4} C. King, ``Maximal p-norms of entanglement breaking channels'',
{\em Quantum Information and Computation}, {\bf 3}, no. 2, 186--190  (2003).

\bibitem{Knew} C. King, reported at the AMS-PTM meeting in Warsaw, Poland,
2 August 2007.

\bibitem{KK}  C. King and N. Koldan    
 ``New multiplicativity results for qubit maps''
 arXiv:quant-ph/0512185

\bibitem{KMNR} 
C. King, K. Matsumoto, M. Nathanson and M.~B. Ruskai,
  ``Properties of Conjugate Channels with Applications to Additivity and Multiplicativity''
    (quant-ph/0509126.
    
 \bibitem{KNR}  C. King, M. Nathanson and M.~B. Ruskai,
  ``Multiplicativity results for entrywise positive maps''
{\em  Lin. Alg. Appl.}  {\bf 404}, 367--379  (2005).
quant-ph/0409181.


\bibitem{KR}  C.King and  M. B. Ruskai,
 ``Comments on multiplicativity of maximal p-norms when p = 2''
  in {\em Quantum Information, Statistics and Probability}  
  ed. by O. Hirota, 102--114 (World Scientific, 2004)   quant-ph/0401026.

   \bibitem{K1} A. Klyachko,  
``Quantum marginal problem and N-representability''
{\em Journal of Physics: Conf. Series}  {\bf 36}, 72--86 (2006).
 quant-ph/0511102
 
\bibitem{LT} E. Lieb and W. Thirring,
``Inequalities for the Moments of the
Eigenvalues of the Schr\"odinger Hamiltonian and Their Relation to Sobolev
Inequalities'', in {\it Studies in Mathematical Physics},  E. Lieb, B.
Simon, A. Wightman eds., pp. 269--303 (Princeton University Press, 1976).
Reprinted in \cite{Lb}

\bibitem{LCV}  Y.-K. Liu, M. Christandl, F. Verstraete 
``N-representability is QMA-complete'' quant-ph/0609125

\bibitem{M} Spyridon Michalakis,
``Multiplicativity of the maximal output 2-norm for depolarized Werner-Holevo channels''
 arXiv:0707.1722

   \bibitem{NR}  M. Nathanson and M.B. Ruskai
``Pauli Diagonal Channels Constant on Axes''
{\em J. Phys. A: Math. Theor.}  {\bf  40},  8171--8204  (2007).
 quant-ph/0611106

\bibitem{Ritt} G. W. Ritter,   
``Quantum Channels and Representation Theory'' 
 {\em J. Math. Phys.} {\bf 46},    (2005)
 (quant-ph/0502153).

\bibitem{R}  D. Ruelle, {\em Statistical Mechanics} (Benjamin, 1969)
Section 2.5.2.

\bibitem{RSW} M. B. Ruskai, S. Szarek, E. Werner, 
``An analysis of completely positive trace-preserving maps $M_2$''
{\em Lin. Alg. Appl.}  {\bf 347}, 159 (2002).

\bibitem{Shir1}  M.E. Shirokov
``The Holevo capacity of infinite dimensional channels and the additivity problem'' 
{\em Commun. Math. Phys.}  {\bf 262}, 137--159  (2006).     quant-ph/0408009


\bibitem{Shir2}  M. E. Shirokov
``The Convex Closure of the Output Entropy of Infinite 
Dimensional Channels and the Additivity Problem''
quant-ph/0608090

\bibitem{ShorEB} P. Shor,
``Additivity of the classical capacity of entanglement-breaking
quantum channels''
{\em  J. Math. Phys.} {\bf 43}, 4334--4340 (2002).


\bibitem{ShorEq} P. W. Shor, 
 ``Equivalence of Additivity Questions in Quantum Information Theory'', 
{\em Commun. Math. Phys.}  {\bf 246}, 453--472 (2004). 
 quant-ph/0305035
 
 \bibitem{ShorPC} P. W. Shor, private communication.   This result
 has been confirmed by M. Nathanson.

 \bibitem{WH} R.~F. Werner and A.~S. Holevo, 
``Counterexample to an additivity conjecture 
for output purity of quantum channels'',
{\em  J. Math. Phys.} {\bf 43},   4353--4357  (2002).

\bibitem{Wp} Andreas Winter,  
``The maximum output p-norm of quantum channels is not multiplicative 
for any $p > 2$''  arXiv:0707.0402

\end{thebibliography}
\end{document}